\documentclass[prl,twocolumn,showpacs,superscriptaddress,floatfix]{revtex4-1}
\usepackage{graphicx,amsfonts,amssymb,amsmath,hyperref}

\newif\ifhyper
\hypertrue
\ifhyper
\hypersetup{
   citecolor = {green},
   colorlinks = {true}, 
   urlcolor = {blue} 
}
\fi

\newcommand{\beq}{\begin{equation}}
\newcommand{\eeq}{\end{equation}}
\newcommand{\beqa}{\begin{eqnarray}}
\newcommand{\eeqa}{\end{eqnarray}}

\def\Longarrow{\protect\@lra}
\def\@lra{\relbar\joinrel\relbar\joinrel\relbar\joinrel%
          \relbar\joinrel\rightarrow}

\begin{document}

\title{Robustness of a perturbed topological phase}

\author{S\'ebastien Dusuel}
\affiliation{Lyc\'ee Saint-Louis, 44 Boulevard Saint-Michel, 75006 Paris, France}

\author{Michael Kamfor}
\affiliation{Lehrstuhl f\"{u}r Theoretische Physik I, Otto-Hahn-Stra\ss e 4, TU Dortmund, 44221 Dortmund, Germany}
\affiliation{Laboratoire de Physique Th\'eorique de la Mati\`ere Condens\'ee, CNRS UMR 7600, \\ 
Universit\'e Pierre et Marie Curie, 4 Place Jussieu, 75252 Paris Cedex 05, France}

\author{Rom\'an Or\'us}
\affiliation{School of Mathematics and Physics, The University of Queensland, QLD 4072, Australia}
\affiliation{Max-Planck-Institut f\"ur Quantenoptik, Hans-Kopfermann-Stra\ss e 1, 85748 Garching, Germany}

\author{Kai Phillip Schmidt}
\affiliation{Lehrstuhl f\"{u}r Theoretische Physik I, Otto-Hahn-Stra\ss e 4, TU Dortmund, 44221 Dortmund, Germany}

\author{Julien Vidal}
\affiliation{Laboratoire de Physique Th\'eorique de la Mati\`ere Condens\'ee, CNRS UMR 7600, \\ 
Universit\'e Pierre et Marie Curie, 4 Place Jussieu, 75252 Paris Cedex 05, France}

\begin{abstract}

We investigate the stability of the topological phase of the toric code model in the presence of a uniform magnetic field by means of variational and high-order series expansion approaches. We find that when this perturbation is strong enough, the system undergoes a topological phase transition whose first- or second-order nature depends on the field orientation. When this transition is of second order, it is in the Ising universality class except for a special line on which the critical exponent driving the closure of the gap varies continuously, unveiling a new topological universality class.
 
\end{abstract}

\pacs{71.10.Pm, 75.10.Jm, 03.65.Vf, 05.30.Pr}

\maketitle

%
%
\emph{Introduction ---}
%
%
The concept of topological quantum order was introduced by Wen in the late 1980s, to characterize the chiral spin state supposed to be relevant for high-temperature superconductivity  \cite{Wen89,*Wen90_1}. 
Since then, it has been shown to be crucial for characterizing different states of matter, among which are fractional quantum Hall states, and it has become the cornerstone of topological quantum computation \cite{Kitaev03,Ogburn99}. 
Topologically ordered quantum systems are mainly characterized by a ground-state degeneracy which depends on the Euler-Poincar\'e characteristic. For connected orientable surfaces, this number is directly related to the genus. Topologically ordered states cannot be characterized by local order parameters and thus fail to be described by Landau symmetry-breaking theory.
Importantly, this nonlocality often implies anyonic statistics and a robustness of the corresponding system with respect to any local perturbation \cite{Kitaev03,Klich10,Bravyi10}, so that they might be used as reliable quantum memories \cite{Dennis02}. 
However, it has early been realized in the seminal paper of Kitaev \cite{Kitaev03} that {\it ``Of course, the perturbation should be small enough, or else a phase transition may occur."}

The main motivation of the present work is precisely to investigate this robustness in the simplest model displaying topological quantum order, namely, the toric code \cite{Kitaev03}, and in the presence of the simplest local perturbation, i.e., a uniform magnetic field. This model, which might be implemented in Josephson junction arrays \cite{Gladchenko09},  may indeed be considered as the ``Ising model of topological quantum phase transitions" and has already been studied for special directions of the field \cite{Hamma08,Trebst07,Tupitsyn10,Vidal09_1,Vidal09_2}  (see also Ref.~\onlinecite{Yu08} for a related problem in Wen's model \cite{Wen03}). 
Here, we address this problem for an arbitrary field direction and determine the extension of the topological phase originating from the zero-field limit.  
To compute this phase diagram, one faces several difficulties since ({\it i}) the lack of a local order parameter prohibits any field-theoretical approach to analyze the critical properties and ({\it ii}) one can neither perform Monte-Carlo simulations (sign problem) nor reliable exact diagonalizations (only small sizes are available). 
Consequently, we combine two different techniques. First, we perform high-order series expansion in the small-field limit using perturbative continuous unitary transformations (PCUT) \cite{Wegner94, *Stein97,*Knetter00_1,*Knetter03_1} and compute the ground-state energy as well as the low-energy gap. Unfortunately, although such an expansion is very efficient to characterize second-order transitions \cite{Vidal09_1}, it cannot locate first-order transitions except in very special situations \cite{Vidal09_2}. 
Second, we use a variational approach based on infinite projected entangled pair states (iPEPS) \cite{Verstraete04_3,Jordan08,Orus09_2} which is, by contrast, especially sensitive to first-order transitions (see, for instance, Ref.~\onlinecite{Orus09_1}). 
Combining these two methods, we determined the boundaries of the topological phase of the toric code model in an arbitrary uniform magnetic field.  The resulting phase diagram displays many interesting features since, depending on the direction of the field, the breakdown of the topological phase may be achieved through a first- or a second-order transition. In the latter case, the universality class is always of Ising type except on a special line where the critical exponent driving the closure of the gap varies continuously.

%
%
\emph{Model and limiting cases---}
%
%
The Hamiltonian of the toric code in a uniform magnetic field reads
%
%
\begin{equation*}
 \label{eq:ham}
 H = - J \sum_{s} A_s  - J \sum_{p} B_p- \boldsymbol{h}\cdot\sum_{i} {\boldsymbol \sigma}_i,
\end{equation*} 
%
%
where $A_s=\prod_{i \in s} \sigma_i^x$ and $B_p =\prod_{i \in p} \sigma_i^z$ ($\sigma_i^\alpha$'s are the usual Pauli matrices).
Subscript $s$ ($p$) refer to sites (plaquettes) of a square lattice and $i$ runs over all bonds where spins  are located \cite{Kitaev03}. Without loss of generality, we restrict our study to $h_\alpha \geqslant 0$, the spectrum being unchanged under the transformation $h_\alpha \rightarrow -h_\alpha$.

In the  zero-field limit, $H$ is exactly solvable since $[A_s,B_p]=0$. As shown in Ref.~\onlinecite{Kitaev03}, the ground-state degeneracy depends on the surface topology so that the system is topologically ordered. In this limit, the ground-state energy per spin is $e_0=-J$. Elementary excitations are obtained by acting onto the ground states with $\sigma_i^z$ (charge excitations) or $\sigma_i^x$ (flux excitations) operators which locally change the eigenvalues of $A_s$ or $B_p$. On a torus, only pairs of such elementary excitations can be created so that, in this case, one has an equidistant spectrum with an energy gap $\Delta=4J$. By contrast, for open boundary conditions, the gap is $\Delta=2J$ since one can create states with only one charge or only one flux. Charges and fluxes behave individually as hard-core bosons but have mutual anyonic (semionic) statistics \cite{Kitaev03}.
In the opposite limit $J=0$, the ground state is unique and fully polarized in the field direction whatever the boundary conditions so it is not topologically ordered anymore. It is thus obvious that at least one phase transition occurs between these two limiting cases. 

In the presence of the field, $A_s$'s and $B_p$'s are no longer conserved so that  $H$ is no longer integrable. However, for some special directions of the field, some mappings onto well-known problems exist. In the following and without loss of generality, we set $J=1/2$. 

%
%
$ \bullet$ $ h_y=0$ --
%
%
 The first simple example is obtained when the field points in the $x$ (or $z$) direction. In this case, the problem is equivalent to the two-dimensional (2D) transverse-field quantum Ising model \cite{Hamma08,Trebst07} which is known to display a second-order transition for \mbox{$h_x=0.1642(2)$}  \cite{He90}.
When both $x$ and $z$ components of the field are nonvanishing, the Hamiltonian $H$ is equivalent to the 3D classical $\mathbb{Z}_2$ gauge Higgs model \cite{Tupitsyn10}. In the plane $h_y=0$, the phase diagram consists of two second-order lines which originate from the Ising points ($h_x= 0$ and $h_z=0$) and  intersect at a multicritical point located at the symmetric point $h_x=h_z=0.1703(2)$ \cite{Vidal09_1}.

%
%
$ \bullet$ $ h_x=h_z=0$ --
%
%
When the field points in the $y$ direction, $H$ is self-dual (its spectrum is invariant under the exchange $h_y \leftrightarrow J$). In addition, it is isospectral to the 2D quantum compass model \cite{Chen07_1} which is also equivalent to that of the Xu-Moore model \cite{Nussinov05_1}. 
In this case, a first-order transition occurs at the  point $h_y=J$ \cite{Orus09_1,Vidal09_2}.

%
%
\emph{Methods~: {\rm PCUT} and {\rm iPEPS} ---}
%
%
Away from these special directions, no mapping onto existing models is known so far. To analyze the full phase diagram, we have first computed the low-energy spectrum using the PCUT (together with the finite-lattice method \cite{Dusuel10_1}) in the small-field limit,  which has already been proven to be very efficient in this context \cite{Vidal09_1,Vidal09_2}. This approach provides a natural description in terms of dressed anyonic quasiparticles in the thermodynamical limit. We focused on the ground-state energy per spin $e_0$ and the 
one-quasiparticle gap $\Delta$ which have been computed at order 10 and 8, respectively.
 The lengthy expressions of these quantities can be found in the supplementary material. We emphasize that, at such high orders, $e_0$  ($\Delta$) is determined with a relative precision lower than $10^{-3}$ ($10^{-2}$) for all directions of the magnetic field and inside the topological phase. Of course, as for any series expansion, such error bars can only be roughly estimated using various resummation schemes (see Ref.~\onlinecite{Oitmaa06} for a detailed discussion).

The PCUT method allows us to determine the set of points $(h_x,h_y,h_z)$ where $\Delta$ vanishes and hence where there might be a continuous transition. However, we know that for $h_x=h_z=0$, the transition is first order and thus not detectable by the condition $\Delta=0$. This is the main reason for using a complementary tool based on a va\-riational approach, the so-called
iPEPS algorithm, which also allows to estimate $e_0$ in the thermodynamic limit with a rather good accuracy \cite{Jordan08,Orus09_2, Orus09_1}. 
The main parameter in this method is the so-called bond dimension $D$ of the PEPS tensors 
\cite{Verstraete04_3,Jordan08,Orus09_2} which drives the amount of entanglement of the ansatz states. 

Our main motivation for choosing such ansatz states is that eigenstates of the toric code (zero-field limit) are described by $D=2$ PEPS \cite{Verstraete06} whereas for $J=0$, eigenstates of $H$ are $D=1$ (completely separable) states. Obviously, in the large $D$ limit, this va\-riational method gives the exact ground state but, in practice, we have checked that the difference between $D=2$ and $D=3$ lies within the error bars of the PCUT calculation so that, for the sake of simplicity, we restrict our analysis to $D=2$ only. Once the bond parameter is fixed, one still has the freedom to choose different ansatz states. Here, we choose a PEPS structure similar to that proposed in Ref.~\onlinecite{Jordan08}, but we allow four different tensors for the four spins of each elementary plaquette (instead of two in Ref.~\onlinecite{Jordan08}). Such a choice leads to $8D^4-1$  variational parameters  (instead of $4D^4-1$) and thus improves the results. 
Other technical details of the algorithm have also been adapted to tackle four-spin interactions. 

One may argue that in order to capture the topological properties of the ground state in the general case (such as a nontrivial topological entropy \cite{Kitaev06_2,*Levin06}), one would need to implement some gauge symmetries in the tensor network ansatz 
\cite{Schuch10,*Chen10_1,*Tagliacozzo10}. But, such properties reflect nonlocal features and are not crucial for computing local quantities such as the ground-state energy. 

\begin{figure}[t]
\includegraphics[width= \columnwidth]{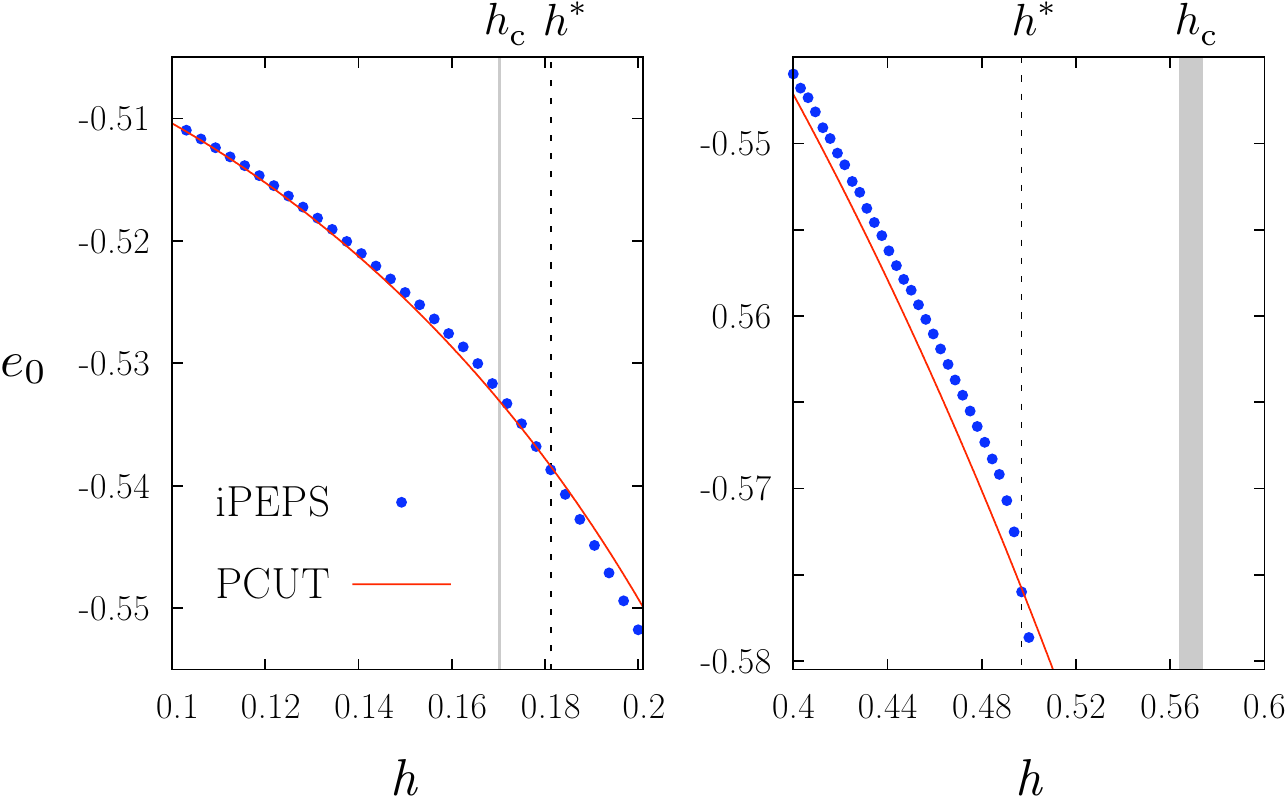}
\caption{(Color online) Comparison of iPEPS and PCUT ground-state energy for two different field directions. The width of the (gray) band defining $h_{\rm c}$ results from different Dlog Pad\'e approximants.
Left~: $\boldsymbol{h}=h (1,0,1)$  and  $h^* > h_{\rm c}$ indicating a second-order transition at $h_{\rm c}$. 
Right~: \mbox{$\boldsymbol{h}=h (\cos \tfrac{7 \pi}{16},\sin\tfrac{7 \pi}{16},\cos\tfrac{7 \pi}{16})$} and  $h^* < h_{\rm c}$  indicating a first-order transition at $h^*$.
}
\label{fig:PEPS}
\end{figure}

Keeping all these approximations in mind, let us describe the general strategy to determine a transition point and its nature (first or second order).
For a fixed direction of the field we wish to compute the critical value of the field's strength $h$ beyond which the system is no more in a topological phase. To do so, one proceeds in three steps~:
%
%
({\it i}) compute the iPEPS ground-state energy $e_0^{\rm iPEPS}$ for different values of $h$ by minimizing the tensor parameters;
({\it ii}) determine the point $h^*$ at which $e_0^{\rm iPEPS}< e_0^{\rm PCUT}$ where $e_0^{\rm PCUT}$ denotes the PCUT ground-state energy;
({\it iii}) compute the value $h_{\rm c}$ for which the one-quasiparticle gap vanishes using the PCUT expression of $\Delta$ and resummation techniques.
%
%
Then two situations must be distinguished.
Either $h^*>h_c$, in which case we can trust the PCUT result and its prediction of a second-order transition at $h_c$. The iPEPS approach is indeed variational and invalidates the PCUT's prediction when $e_0^{\rm iPEPS}< e_0^{\rm PCUT}$.
Or $h^*<h_c$, in which case a transition occurs before the gap $\Delta$ vanishes. This means that there are some level 
crossings due to higher-energy levels which are not captured by the PCUT approach, indicating a first-order transition confirmed by the discontinuity of the slope of the iPEPS energy [see {\it e.g.} Fig.~\ref{fig:PEPS} (right)]. 
Note that one may indeed directly compute the derivative of $e_0^{\rm iPEPS}$ as a function of $h$ and look for singularities but this approach is less precise. 
Obviously, the precision in the determination of $h^*$ and $h_{\rm c}$ plays a fundamental role in this scheme.  For a given direction, the maximum orders at which we computed $e_0$ and $\Delta$ as well as the form of the chosen variational states allow us to estimate the transition point with an accuracy of a few percent as can be seen in Fig.~\ref{fig:PEPS}. 
%
%
%
%

\begin{figure}[t]
\includegraphics[width=\columnwidth]{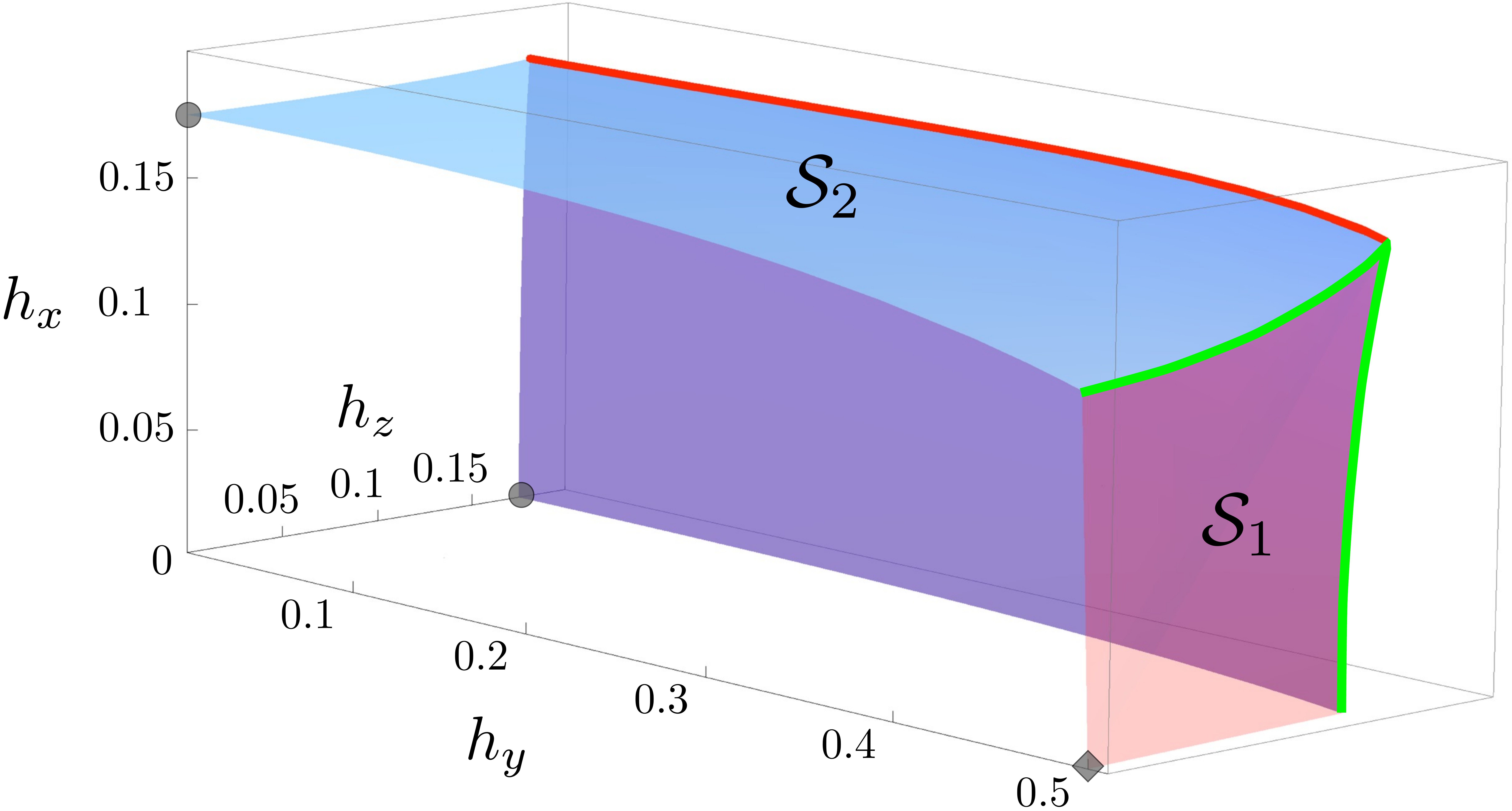}
\caption{(Color online) Sketch of the 3D phase diagram. Dots correspond to Ising points and the diamond is the self-dual point of the $h_y$ line. Green lines are the intersections of  the first-order sheet ${\mathcal S}_1$ and the second-order sheet ${\mathcal S}_2$ (computed from the bare series given in supplementary material). The multicritical line $h_x=h_z$ with continuously varying critical exponents is shown as a thick (red) line.}
\label{fig:diagram}
\end{figure}

%
%
\emph{Phase diagram ---}
%
%
A sketch of the 3D phase diagram is shown in Fig.~\ref{fig:diagram} and can be summarized as follows. First, we find that the transition point \mbox{$\boldsymbol{h}=(0,1/2,0)$} is part of a 2D first-order transition sheet 
${\mathcal S}_1$. 
Second, the second-order transition lines of the \mbox{$h_y=0$} plane give rise to a 2D second-order transition sheet ${\mathcal S}_2$ (defined by \mbox{$\Delta=0$}) when the $y$-component of the field is nonvanishing. These sheets that intersect on a nontrivial line define the boundaries of the topological phase. 
Given the difficulty for investigating the full 3D space with iPEPS, we focused on some special planes in which we determined the coordinates of the intersection point of ${\mathcal S}_1$ and ${\mathcal S}_2$. 
For instance, in the $(0,h_y,h_z)$ plane, we found that this intersection occurs around the point $\boldsymbol{h}=(0,0.49,0.11)$. 
When the transition is second order, the gap is expected to behave  as $\Delta\sim (h-h_{\rm c})^{z \nu}$ in the vicinity of the critical point $h_{\rm c}$. Note that here, we do not have access to the dynamical exponent $z$ and to the correlation length exponent $\nu$ independently but only to their product. 
For all investigated directions, we found that $z \nu$ was compatible with the 
well-established Ising value $z \nu=0.630(1)$. 
This leads us to conclude that ${\mathcal S}_2$ lies in the Ising universality class (as was already found in the plane $h_y=0$ \cite{Vidal09_1,Tupitsyn10}) for all directions except for the special case $h_x=h_z$.

%
%
%

\begin{figure}
\includegraphics[width=\columnwidth]{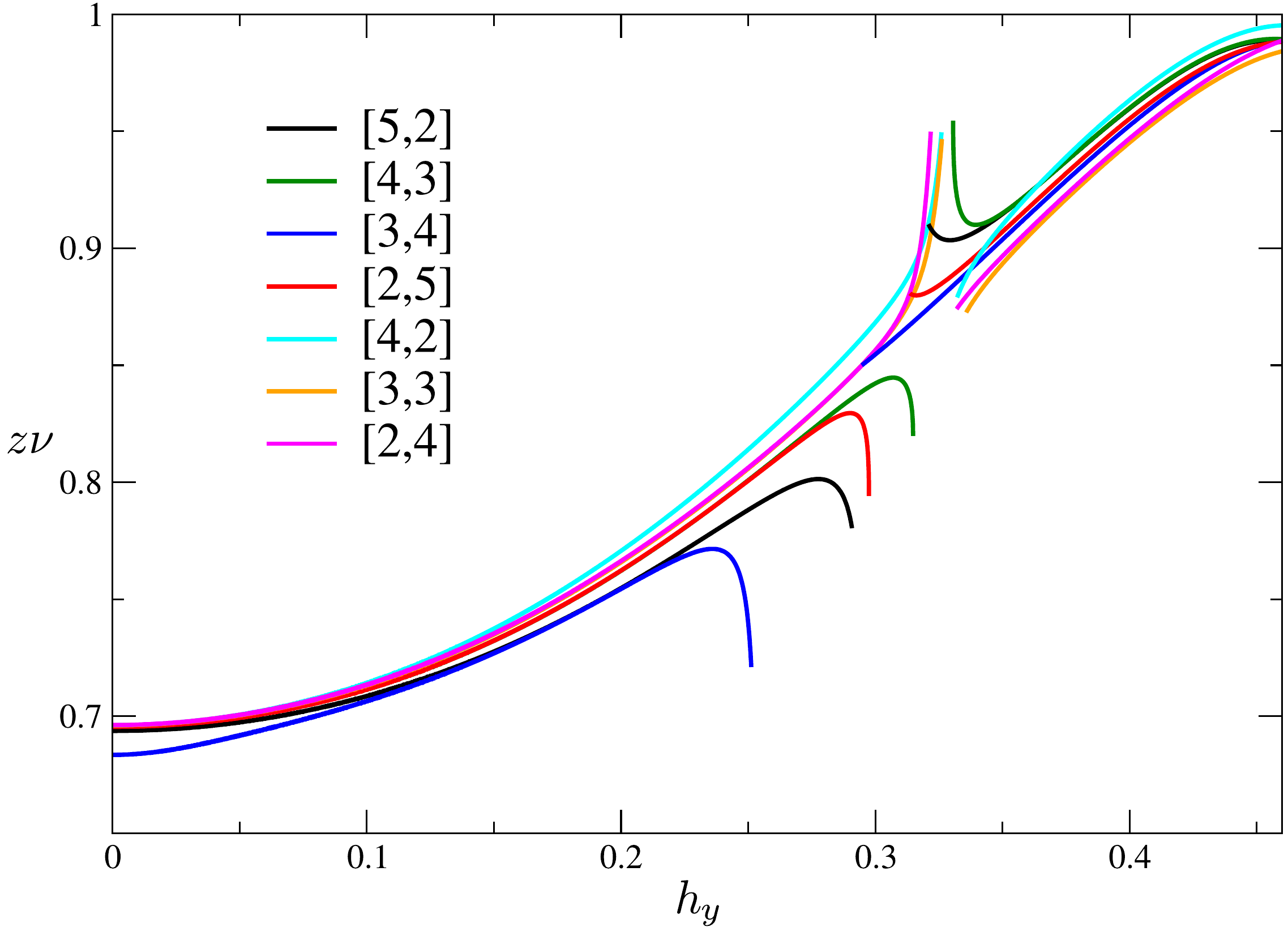}
\caption{(Color online) Critical exponent $z \nu$ as a function of $h_y$ along the line $h_x=h_z$ computed for various Dlog Pad\'e  approximants $[m,n]$. Strange behaviors near $h_y\simeq0.3$ are likely due to spurious pole structures and should not be considered as relevant.}
\label{fig:exponent}
\end{figure}

%
%
\emph{The multicritical line ---}
%
%
As discussed in \cite{Vidal09_1,Tupitsyn10} for $h_y=0$, the two second-order transition lines merge in a multicritical point at $h_x=h_z$ for which the gap exponent  is clearly different from the Ising value. The most important result of the present study is that when $h_y \neq 0$, this multicritical point gives rise to a multicritical line on which this exponent varies continuously. 
First of all, let us point out that the multicritical line intersects ${\mathcal S}_1$ around the point $\boldsymbol{h}=(0.17,0.46,0.17)$. Once again these values are obtained with a relative precision of a few percent. Along this multicritical line, we have computed the exponent $z \nu$ using standard resummation techniques based on Dlog Pad\'e approximants (see Ref.~\onlinecite{Oitmaa06} for details). Our results are displayed in Fig.~\ref{fig:exponent} and show that this exponent varies from 0.69 at $h_y=0$ up to a value close to 1 at $h_y=0.46$ along this line. Except in the range $h_y\in[0.20,0.35]$, one gets a rather good convergence suggesting that divergencies observed in this region are due to spurious poles in the Dlog Pad\'e approximants.
We thus conjecture that $z \nu$ varies continuously and that its variation of $\sim 50\%$ cannot be attributed to extrapolation errors and reveals a new universality class. Since it is not associated to a symmetry breaking but rather reflects the breakdown of a topological phase, we will call it topological. 

At this stage, it is difficult to determine the key ingredients for a system to belong to this class (since we do not have any local order parameter) but it is likely that the mutual semionic statistics of charges and fluxes is one of them.  More generally, let us underline that continuously varying critical exponents are not common in two-dimensional quantum systems. During the completion of this work, some conformal quantum critical lines in 2+1 dimensions have been proposed \cite{Ardonne04,Isakov11} but their relevance for the toric code in a magnetic field is still an open question. 
%
%
%
%

%
%
\emph{Discussion and outlook ---}
%
%
In the present work, we have determined the boundaries of the topological phase of the toric code in a field  using two state-of-the-art and complementary methods. This topological ``bubble" is made of first-order and second-order sheets. Interestingly, second-order transitions seem to be in the Ising universality class except on a multicritical line on which the gap vanishes with continuously varying exponents giving rise to a new ``topological" universality class. 
Of course, it would also be valuable to study the large-field limit of this model to investigate the outer part of the bubble. Notably the fate of the first-order line observed in the $h_y=0$ plane \cite{Vidal09_1,Tupitsyn10} is an interesting question. Finally, a complete understanding of  the low-energy spectrum of the topological phase certainly requires the study of bound states as already seen in the transverse-field case \cite{Vidal09_2}.\\

R.O. acknowledges financial support from the ARC, UQ, and the EU through a Marie Curie International Incoming Fellowship, as well as discussions with B. Bauer, P. Corboz, J. Jordan, L. Tagliacozzo, and G. Vidal. K.P.S. and M.K. acknowledge financial support from the DFG and thank ESF and EuroHorcs for funding through the EURYI.

%

\onecolumngrid

\section*{Supplementary material}

Here are the series expansions obtained using the PCUT method in the small-field limit $h_x,h_y,h_z \ll J $. Setting $S_k=h_x^k+h_z^k$, $P_{2k}=h_x^k h_z^k$ and $J=1/2$, the ground-state energy per spin $e_0$ at order 10 reads
%
%
\begin{eqnarray}
 e_0&=& - \frac{1}{2} - \frac{S_2}{2}- \frac{h_y^2}{4}  -\frac{15 S_4}{8}-\frac{7 S_2 h_y^2}{32} +\frac{P_4}{4}- \frac{13 h_y^4}{192}   - \frac{147 S_6}{8}- \frac{371 S_4 h_y^2}{128} +\frac{113 S_2 P_4}{32} - \frac{1045 S_2 h_y^4}{3456}   + \frac{2003 P_4 h_y^2}{384}  \nonumber \\
 &&    - \frac{197 h_y^6}{3072} - \frac{18003 S_8}{64}  - \frac{1954879 S_6 h_y^2}{36864} +  \frac{6685 S_4 P_4}{128}- 
 \frac{34054175 S_4 h_y^4}{3981312} +\frac{146861 S_2 P_4 h_y^2}{2304} - \frac{15343549 S_2 h_y^6}{26542080} \nonumber \\
 &&+ \frac{20869 P_8}{384} +\frac{5020085  P_4 h_y^4}{497664} - \frac{163885 h_y^8}{1769472} 
 -\frac{5420775 S_{10}}{1024}  -\frac{1563459523 S_{8}h_y^2}{1327104} +\frac{39524033 S_{6}P_4}{36864}\nonumber \\
 && -\frac{1115105409427 S_{6}h_y^4}{5733089280} +\frac{10058235445 S_{4} P_4 h_y^2}{7962624}-\frac{4219640835497 S_{4}h_y^6}{191102976000}+\frac{5650925 S_{2}P_8}{6912}+\frac{20854097563 S_{2} P_4 h_y^4}{143327232}  \nonumber \\
 &&-\frac{483890940281 S_{2}h_y^8}{382205952000}
 +\frac{1202498305 P_{8}h_y^2}{1990656}+
 \frac{1994817656221 P_{4}h_y^6}{71663616000} -\frac{186734746441 h_y^{10}}{1146617856000}.
 \end{eqnarray}
%
%
Similarly, for $0\leq h_x \leq h_z$, and the one-quasiparticle (dressed charge) gap $\Delta$ at order 8 reads
%
%
\begin{eqnarray}
\Delta&=& 1-4 h_z -h_y^2-4 h_z^2+2 h_x^2 h_z+\frac{11}{4} h_y^2 h_z-12 h_z^3  + 5 h_x^4 +17 h_x^2 h_y^2 - \frac{15}{16} h_y^4+3 h_x^2 h_z^2 -9 h_y^2 h_z^2-36 h_z^4 +\frac{27}{2}h_x^4 h_z \nonumber \\
 && +\frac{17}{4} h_y^4 h_z+\frac{9}{4} h_x^2 h_y^2 h_z+\frac{83}{4} h_x^2 h_z^3 +\frac{473}{64} h_y^2 h_z^3-176 h_z^5 +92 h_x^6+\frac{14267}{96} h_y^2 h_x^4+71 h_z^2 h_x^4 +\frac{13621}{1152} h_y^4 h_x^2+ 63 h_z^4 h_x^2 \nonumber \\
 &&+\frac{1305}{8} h_y^2 h_z^2  h_x^2-\frac{575}{384} h_y^6 - \frac{2625}{4}h_z^6-\frac{7971}{64} h_y^2 h_z^4-\frac{135619}{3456}h_y^4 h_z^2 +\frac{495}{2} h_x^6 h_z +\frac{1142149}{4608} h_x^4 h_y^2 h_z-\frac{3031}{13824} h_x^2 h_y^4 h_z\nonumber \\
 &&+\frac{799973}{110592} h_y^6 h_z +\frac{925}{4} h_x^4 h_z^3+\frac{13807}{48} h_x^2 h_y^2 h_z^3+\frac{1782929}{20736} h_y^4 h_z^3+ \frac{28633}{64} h_x^2 h_z^5-\frac{238621}{1152} h_y^2 h_z^5-\frac{14771}{4} h_z^7+\frac{35649}{16} h_x^8\nonumber \\
 && +\frac{7715431}{3072} h_x^6 h_y^2+\frac{3032191}{31104} h_x^4 h_y^4+\frac{98263727}{3981312} h_x^2 h_y^6- \frac{26492351}{7962624} h_y^8+\frac{80999}{96} h_x^6 h_z^2+\frac{2199571}{4608} h_x^4 h_y^2 h_z^2\nonumber \\
 &&+ \frac{24547709}{165888} h_x^2 h_y^4 h_z^2-\frac{1495320677}{19906560} h_y^6 h_z^2+ \frac{19263}{16} h_x^4 h_z^4+ \frac{5186533}{1728} h_x^2 h_y^2 h_z^4- \frac{1760584999}{1990656} h_y^4 h_z^4+ \frac{118029}{64} h_x^2 h_z^6\nonumber \\
 &&-\frac{4663837}{1728} h_y^2 h_z^6- \frac{940739}{64} h_z^8.
\end{eqnarray}
%
%

The (dressed flux) gap for $h_x > h_z$ is straightforwardly obtained by exchanging $h_x$ and $h_z$ in this expression.\\\\

\underline{\it Errata}~:\\

- For $h_y=0$, one recovers expressions given in Eq.~(8) of Ref.~\onlinecite{Vidal09_1} up to a typo~: the term proportional to $(h_x^8+h_z^8)$ is missing. \\

- For $h_x=h_z=0$, one recovers expressions given in Eq.~(4)  of Ref.~\onlinecite{Vidal09_2} up to a typo~: the term proportional to $t^{10}$ must be corrected by a factor $2$. 

\end{document}